\documentclass[eqsecnum,superscriptaddress,nofootinbib,11pt]{revtex4}
\usepackage{amsmath}
   \usepackage{bm}
\usepackage{amsthm,amscd}
\usepackage{mathrsfs}
\usepackage{verbatim}
\usepackage{amsfonts,amsmath,latexsym,amssymb,txfonts,bm}
\usepackage[american]{babel}
\usepackage{dsfont}
\usepackage[dvips]{graphicx}
\usepackage{latexsym}
\usepackage{epsfig}
\newcommand{\diff}{\mathrm{d}}

\def\beq{\begin{eqnarray}}
\def\eeq{\end{eqnarray}}

\begin{document}
\title{The Casimir Effect for Thick Pistons}

\author{Guglielmo Fucci\footnote{Electronic address: fuccig@ecu.edu}}
\affiliation{Department of Mathematics, East Carolina University, Greenville, NC 27858 USA}

\date{\today}
\vspace{2cm}
\begin{abstract}

In this work we analyze the Casimir energy and force for a {\it thick} piston configuration. This study is performed
by utilizing the spectral zeta function regularization method. The results we obtain for the Casimir energy and force depend explicitly 
on the parameters that describe the general self-adjoint boundary conditions imposed. Numerical results for the Casimir force are 
provided for specific types of boundary conditions and are also compared to the corresponding force on an infinitely thin piston.

\end{abstract}
\maketitle

\section{Introduction}

Variations in the vacuum structure of a quantum field give rise to what is commonly know as the Casimir effect \cite{bordag09,milton01,milo94b,most97b}. 
It its simplest form it manifests itself as a net force between neutral objects \cite{casimir48}. Any attempt at computing the Casimir energy and force 
for almost all geometric configurations of neutral objects necessarily leads to divergences that need to be regularized 
through the use of appropriate formal methods \cite{bordag09}. 
A particular example of a configuration for which the Casimir force is, generally, well defined is provided by the so-called Casimir piston. 
The most general Casimir piston can be described as consisting of two compact manifolds having a common boundary which is referred to as the piston.
Casimir piston configurations were first introduced in \cite{caval04}, and since then they have been the subject of intense research 
(see e.g. \cite{barton06,edery06,edery07,edery08,hertz05,hertz07,kirsten09,li97,marachevsky07}). 
The main reason for such popularity lies in the fact that while the Casimir energy might present divergences, the corresponding 
Casimir force on the piston is well defined. It is important to point out, however, that this does not hold, 
in general, when the piston configuration has a non vanishing curvature \cite{fucci12}.

Most of the research that has been performed on the Casimir effect for piston configurations has been focused almost excursively on ``idealized" cases.
These are cases characterized by Dirichlet, Neumann or hybrid (mixed) boundary conditions imposed at the boundary of the piston configuration, and on the piston itself.
Moreover, in idealized cases, the piston is always assumed to be infinitely thin. It is not very difficult to realize that the results obtained 
for the Casimir energy and force in idealized cases are unsuitable for the description of piston configurations consisting of real materials for two main reasons:
1) Physical properties of real materials might not be appropriately described by any of the ideal boundary conditions (Dirichlet, Neumann or hybrid) 
and, 2) real materials do have a finite thickness. An attempt at addressing the first problem can be found for 
instance in \cite{fucci15} where the Casimir effect is studied for a piston configuration endowed with general boundary conditions. 
A different approach was undertaken in \cite{acto95,bea13,fucci11,eliz97} where the ideal boundary conditions 
were replaced by a suitable potential function which would better describe the physical properties of real materials.

The Casimir effect for materials of finite thickness, instead, has been analyzed within the framework of piston configurations for instance in \cite{teo10,barton06}.
The main purpose of this work is to provide a more comprehensive study of the Casimir effect for piston configurations in which we consider general boundary conditions
and a piston of finite thickness. 
It is worth noting that in this approach the boundary conditions imposed on one side of the thick piston are allowed to be independent of
the boundary conditions imposed on the other side. This specific setup could be very useful in describing 
pistons consisting of certain anisotropic materials (those for which the anisotropy occurs along the thickness of the piston). 

In this paper we will use the spectral zeta function regularization technique in order to analyze the Casimir energy and force associated with a 
thick piston. The spectral zeta function, $\zeta(s)$, of the problem under consideration can be used to find the Casimir energy according to the 
formula 
\cite{bordag09,bytsenko03,elizalde94,elizalde,kirsten01}
\begin{equation}\label{0}
E_{\textrm{Cas}}=\lim_{\varepsilon\to 0}\frac{\mu^{2\varepsilon}}{2}\zeta\left(\varepsilon-\frac{1}{2}\right)\;,
\end{equation}   
with $\mu$ representing a parameter with the dimensions of a mass. Due to the fact that the spectral zeta function 
generally develops a simple pole at the point $s=-1/2$ \cite{kirsten01}, the Casimir energy presents divergences which are easily shown in the formula
\begin{equation}\label{0a}
E_{\textrm{Cas}}=\frac{1}{2}\textrm{FP}\,\zeta\left(-\frac{1}{2}\right)+\frac{1}{2}\left(\frac{1}{\varepsilon}+\ln\mu^{2}\right)\textrm{Res}\,\zeta\left(-\frac{1}{2}\right)+O(\varepsilon)\;,
\end{equation}
with $\textrm{FP}$ and $\textrm{Res}$ denoting, respectively, the finite part and the residue. It is clear, from the previous expression, 
that the Casimir energy becomes a well defined quantity when the residue of the spectral zeta function at $s=-1/2$ vanishes. 
In the case of piston configurations, the Casimir energy depends explicitly
on the position of the piston $a$, and the corresponding Casimir force is obtained as follows
\begin{equation}\label{0b}
F_{\textrm{Cas}}(a)=-\frac{\partial}{\partial a}E_{\textrm{Cas}}(a)\;.
\end{equation}
In order to get a Casimir force on the piston devoid of meaningless divergences the residue of the zeta function appearing in (\ref{0a})
has to be independent of the position of the piston $a$.

The outline of the paper is as follows. In the next Section we describe in details the thick piston configuration and obtain the associated spectral zeta function. 
In Section \ref{sec3} we perform the analytic continuation of the spectral zeta function and obtain an explicit expression for the Casimir energy of the form indicated 
in (\ref{0a}). Plots describing how the Casimir force on a thick piston varies with respect to the position of the piston itself and its thickness 
are provided in Section \ref{sec4} for several specific examples of boundary conditions. 
The conclusions summarize our main results and point out further directions of research.

\section{Spectral zeta function for the piston}\label{sec2}

In order to describe the geometric configuration associated with a thick piston we consider a $D$-dimensional product manifold of the type 
$M=[0,L]\times N$ where $[0,L]\subset \mathbb{R}$ and $N$ is a $(D-1)$-dimensional compact Riemannian manifold with or without boundary.  
Let $a$ denote an arbitrary point in the interval $(0,L)$, and let $\epsilon>0$. We then define the manifolds $N_{a-\epsilon/2}$ and
$N_{a+\epsilon/2}$ to be the cross-sections of $M$ at the point $a-\epsilon/2$ and $a+\epsilon/2$, respectively. The 
cross-sections constructed above naturally divide the manifold $M$ in three regions $M_{I}$, $M_{II}$, and $M_{III}$. 
The regions $M_{I}=[0,a-\epsilon/2)\times M$ and $M_{III}=(a+\epsilon/2,L]\times M$ will be referred to as the left, respectively right, chamber of the piston configuration 
while the region $M_{II}=[a-\epsilon/2,a+\epsilon/2]\times M$ will describe the piston itself of thickness $\epsilon$. It is clear that 
in order to have a proper piston configuration we need to assume that $\epsilon\in[0,L)$.
It is not very difficult to realize that in the limit $\epsilon\to 0$ the thick piston outlined above reduces to the familiar infinitely thin Casimir piston
with a general cross-section (see for instance \cite{fucci15}).

In this work we focus our attention on a massless scalar field confined on the piston $M$. The propagation of the scalar field in each region of the piston configuration 
is described by the following differential equation
\begin{equation}\label{1}
\left(-\frac{\diff^{2}}{\diff x^{2}}-\Delta_{N}\right)\phi_{j}=\alpha_{j}^{2}\phi_{j}\;,
\end{equation}    
with $\Delta_{N}$ representing the Laplace operator on the base manifold $N$ and $j=\{I, II,III\}$. By using separation of variables the general solution
to (\ref{1}) can be written as a product $\phi_{j}=\varphi_{j}(\lambda,x)\Phi({\bf x})$ where the function $\Phi({\bf x})$ solves the 
equation $-\Delta_{N}\Phi({\bf x})=\lambda^{2}\Phi({\bf x})$ while the functions $\varphi_{j}(\lambda,x)$ are solutions to the second-order 
differential equation 
\begin{equation}\label{2}
\left(-\frac{\diff^{2}}{\diff x^{2}}+\lambda^{2}\right)\varphi_{j}(\lambda,x)=\alpha^{2}_{j}\varphi_{j}(\lambda,x)\;.
\end{equation}
The spectral zeta function needed for the analysis of the Casimir energy associated with the thick piston configuration is written as the sum
\begin{equation}\label{3}
\zeta(s)=\zeta_{I}(s)+\zeta_{II}(s)+\zeta_{III}(s)\;,
\end{equation} 
where $\zeta_{j}(s)$ represents the spectral zeta function of region $M_{j}$ and is defined as
\begin{equation}\label{4}
\zeta_{j}(s)=\sum_{\alpha_{j}}\alpha_{j}^{-2s}\;.
\end{equation}  
It is well known \cite{kirsten01}, from the general theory of spectral zeta function, that $\zeta_{j}(s)$ in (\ref{4}), and 
consequently $\zeta(s)$ in (\ref{3}), are analytic functions in the region of the complex plane $\Re(s)>D/2$. 

As we have already mentioned earlier, in order to obtain the Casimir energy for our piston configuration we need to analyze the 
behavior of the zeta function in the neighborhood of the point $s=-1/2$. This point, however, lies outside of the region of convergence $\Re(s)>D/2$. This last remark 
implies that it is necessary to perform the analytic continuation of $\zeta(s)$ to the region $\Re(s)\leq D/2$ containing $s=-1/2$.
To perform the desired analytic continuation we utilize a suitable integral representation of the zeta functions $\zeta_{j}(s)$ \cite{kirsten01}. A key 
ingredient of this representation is a function which provides an implicit equation for the eigenvalues $\alpha_{j}$. Such function can be easily found by 
making use of the boundary conditions. Here we will impose, on the differential equation (\ref{2}), general separated boundary conditions that lead to a self-adjoint
boundary value problem \cite{fucci15,fucci15a}. These boundary conditions are of the form \cite{zettl}
\begin{eqnarray}\label{5}
A_{1}\varphi_{I}(\lambda,0)+A_{2}\varphi^{\prime}_{I}(\lambda,0)&=&0\;,\nonumber\\
B_{1}\varphi_{I}\left(\lambda,a-\frac{\epsilon}{2}\right)-B_{2}\varphi^{\prime}_{I}\left(\lambda,a-\frac{\epsilon}{2}\right)&=&0\;,
\end{eqnarray}
in region $I$,
\begin{eqnarray}\label{6}
B_{1}\varphi_{II}\left(\lambda,a-\frac{\epsilon}{2}\right)-B_{2}\varphi^{\prime}_{II}\left(\lambda,a-\frac{\epsilon}{2}\right)&=&0\;,\nonumber\\
C_{1}\varphi_{II}\left(\lambda,a+\frac{\epsilon}{2}\right)+C_{2}\varphi^{\prime}_{II}\left(\lambda,a+\frac{\epsilon}{2}\right)&=&0\;,
\end{eqnarray}
in region $II$, and 
\begin{eqnarray}\label{7}
C_{1}\varphi_{III}\left(\lambda,a+\frac{\epsilon}{2}\right)+C_{2}\varphi^{\prime}_{III}\left(\lambda,a+\frac{\epsilon}{2}\right)&=&0\;,\nonumber\\
D_{1}\varphi_{III}\left(\lambda,L\right)-D_{2}\varphi^{\prime}_{III}\left(\lambda,L\right)&=&0\;,
\end{eqnarray}
in region $III$ where $A_{1},A_{2},B_{1},B_{2},C_{1},C_{2}, D_{1}, D_{2}\in\mathbb{R}$ satisfying 
the conditions $(A_{1},A_{2})\neq(0,0)$, $(B_{1},B_{2})\neq (0,0)$, $(C_{1},C_{2})\neq (0,0)$,
and $(D_{1},D_{2})\neq (0,0)$.  

In all three regions the general solution of the differential equation (\ref{2}) is of the form
\begin{equation}\label{8}
\varphi_{j}(x)=c_{j}^{(+)}\exp\left\{i\sqrt{\alpha_{j}^{2}-\lambda^{2}}(x-x_{j})\right\}+c_{j}^{(-)}\exp\left\{-i\sqrt{\alpha_{j}^{2}-\lambda^{2}}(x-x_{j})\right\}\;,
\end{equation}
where we have used the notation $x_{I}=0$, $x_{II}=a-\epsilon/2$, and $x_{III}=a+\epsilon/2$. Now, we choose  
the coefficients $c_{j}^{(+)}$ and $c_{j}^{(-)}$ so that the boundary condition at the left boundary in each region is automatically satisfied. 
In doing so, one obtains
\begin{eqnarray}\label{9}
\varphi_{I}(\lambda,0)=-A_{2}\;&,&\quad \varphi^{\prime}_{I}(\lambda,0)=A_{1}\;,\\
\varphi_{II}\left(\lambda,a-\frac{\epsilon}{2}\right)=B_{2}\;&,&\quad \varphi^{\prime}_{II}\left(\lambda,a-\frac{\epsilon}{2}\right)=B_{1}\;,\\
\varphi_{III}\left(\lambda,a+\frac{\epsilon}{2}\right)=-C_{2}\;&,&\quad \varphi^{\prime}_{III}\left(\lambda,a+\frac{\epsilon}{2}\right)=C_{1}\;.
\end{eqnarray}
The above relations allow us to rewrite the solution $\varphi_{j}(x)$ as
\begin{equation}\label{10}
\varphi_{j}(x)=\frac{\varphi_{j}^{\prime}(\lambda,x_{j})}{\sqrt{\alpha_{j}^{2}-\lambda^{2}}}\sin\left[\sqrt{\alpha_{j}^{2}-\lambda^{2}}(x-x_{j})\right]
+\varphi_{j}(\lambda,x_{j})\cos\left[\sqrt{\alpha_{j}^{2}-\lambda^{2}}(x-x_{j})\right]\;.
\end{equation}
By imposing the remaining boundary conditions on the right boundary of each region we obtain equations that implicitly provide the 
eigenvalues $\alpha_{j}$. More precisely, in region $I$ we have
\begin{eqnarray}\label{11}
f^{I}_{\lambda}\left(\alpha,a-\frac{\epsilon}{2}\right)&=&\left(\frac{A_{1}B_{1}}{\sqrt{\alpha_{I}^{2}-\lambda^{2}}}-A_{2}B_{2}\sqrt{\alpha_{I}^{2}-\lambda^{2}}\right)\sin\left[\sqrt{\alpha_{I}^{2}-\lambda^{2}}\left(a-\frac{\epsilon}{2}\right) \right]\nonumber\\
&-&\left(A_{2}B_{1}+A_{1}B_{2}\right)\cos\left[\sqrt{\alpha_{I}^{2}-\lambda^{2}}\left(a-\frac{\epsilon}{2}\right) \right]=0\;,
\end{eqnarray}  
in region $II$ we have
\begin{eqnarray}\label{12}
f^{II}_{\lambda}\left(\alpha,\epsilon\right)&=&\left(\frac{C_{1}B_{1}}{\sqrt{\alpha_{II}^{2}-\lambda^{2}}}-C_{2}B_{2}\sqrt{\alpha_{II}^{2}-\lambda^{2}}\right)\sin\left(\sqrt{\alpha_{II}^{2}-\lambda^{2}}\epsilon \right)\nonumber\\
&+&\left(C_{2}B_{1}+C_{1}B_{2}\right)\cos\left(\sqrt{\alpha_{I}^{2}-\lambda^{2}}\epsilon \right)=0\;,
\end{eqnarray}
and, finally, in region $III$ we obtain 
\begin{eqnarray}\label{13}
f^{III}_{\lambda}\left(\alpha,L-a-\frac{\epsilon}{2}\right)&=&\left(\frac{C_{1}D_{1}}{\sqrt{\alpha_{III}^{2}-\lambda^{2}}}-C_{2}D_{2}\sqrt{\alpha_{III}^{2}-\lambda^{2}}\right)\sin\left[\sqrt{\alpha_{III}^{2}-\lambda^{2}}\left(L-a-\frac{\epsilon}{2}\right) \right]\nonumber\\
&-&\left(D_{2}C_{1}+D_{1}C_{2}\right)\cos\left[\sqrt{\alpha_{III}^{2}-\lambda^{2}}\left(L-a-\frac{\epsilon}{2}\right) \right]=0\;.
\end{eqnarray} 

The equations displayed above have simple zeroes which, in general, can be both real and purely imaginary \cite{romeo02,teo09}.
Since, here, we work under the assumption that all the eigenvalues $\alpha_{j}$ are positive, we will only consider instances in which all the solutions of 
the equations (\ref{11}), (\ref{12}), and (\ref{13}) are real. These occur only for values of the parameters in the boundary conditions belonging to 
specific intervals of the real line (see e.g. \cite{fucci15}). To find these intervals it is actually more convenient to consider the 
equations in (\ref{11}), (\ref{12}), and (\ref{13}) with the variable $\alpha_{j}$ replaced with $i\alpha_{j}$, namely
\begin{eqnarray}\label{14}
f^{I}_{\lambda}\left(i\alpha,a-\frac{\epsilon}{2}\right)&=&\left(\frac{A_{1}B_{1}}{\sqrt{\alpha_{I}^{2}+\lambda^{2}}}+A_{2}B_{2}\sqrt{\alpha_{I}^{2}+\lambda^{2}}\right)\sinh\left[\sqrt{\alpha_{I}^{2}+\lambda^{2}}\left(a-\frac{\epsilon}{2}\right) \right]\nonumber\\
&-&\left(A_{2}B_{1}+A_{1}B_{2}\right)\cosh\left[\sqrt{\alpha_{I}^{2}+\lambda^{2}}\left(a-\frac{\epsilon}{2}\right) \right]=0\;,
\end{eqnarray}  
\begin{eqnarray}\label{15}
f^{II}_{\lambda}\left(i\alpha,\epsilon\right)&=&\left(\frac{C_{1}B_{1}}{\sqrt{\alpha_{II}^{2}+\lambda^{2}}}+C_{2}B_{2}\sqrt{\alpha_{II}^{2}+\lambda^{2}}\right)\sinh\left(\sqrt{\alpha_{II}^{2}+\lambda^{2}}\epsilon \right)\nonumber\\
&+&\left(C_{2}B_{1}+C_{1}B_{2}\right)\cosh\left(\sqrt{\alpha_{I}^{2}+\lambda^{2}}\epsilon \right)=0\;,
\end{eqnarray}
and
\begin{eqnarray}\label{16}
f^{III}_{\lambda}\left(i\alpha,L-a-\frac{\epsilon}{2}\right)&=&\left(\frac{C_{1}D_{1}}{\sqrt{\alpha_{III}^{2}+\lambda^{2}}}+C_{2}D_{2}\sqrt{\alpha_{III}^{2}+\lambda^{2}}\right)\sinh\left[\sqrt{\alpha_{III}^{2}+\lambda^{2}}\left(L-a-\frac{\epsilon}{2}\right) \right]\nonumber\\
&-&\left(D_{2}C_{1}+D_{1}C_{2}\right)\cosh\left[\sqrt{\alpha_{III}^{2}+\lambda^{2}}\left(L-a-\frac{\epsilon}{2}\right) \right]=0\;.
\end{eqnarray}
   
Now, it is not very difficult to show that (\ref{14}) has no real zeroes, and hence (\ref{11}) has no purely imaginary solutions,
if the following conditions are satisfied    
\begin{equation}\label{17}
\left\{\frac{A_{2}B_{2}}{(a-\epsilon/2)^{2}A_{1}B_{1}}\leq 0\;,\frac{1}{a-\epsilon/2}\left(\frac{A_{2}}{A_{1}}+\frac{B_{2}}{B_{1}}\right)\geq 1\right\}\;,\quad \textrm{or}\quad
\left\{\frac{A_{2}}{(a-\epsilon/2) A_{1}}\leq 0\;, \frac{B_{2}}{(a-\epsilon/2) B_{1}}\leq 0\right\}\;, 
\end{equation}
for $A_{1}B_{1}\neq0$,
\begin{equation}\label{17a}
\frac{B_{2}}{(a-\epsilon/2) B_{1}}\leq 0\;,\quad A_{1}=0\;,\quad\textrm{and}\quad \frac{A_{2}}{(a-\epsilon/2) A_{1}}\leq 0\;,\quad B_{1}=0\;,
\end{equation}   
and for $A_{1}=B_{1}=0$.
Similar conditions apply to (\ref{16}). In fact, (\ref{13}) has no purely imaginary zeroes when
\begin{eqnarray}\label{18}
\lefteqn{\left\{\frac{C_{2}D_{2}}{(L-a-\epsilon/2)^{2}C_{1}D_{1}}\leq 0\;,\frac{1}{L-a-\epsilon/2}\left(\frac{C_{2}}{C_{1}}+\frac{D_{2}}{D_{1}}\right)\geq 1\right\}\;,\quad \textrm{or}}\nonumber\\
&&\left\{\frac{C_{2}}{(L-a-\epsilon/2) C_{1}}\leq 0\;, \frac{D_{2}}{(L-a-\epsilon/2) D_{1}}\leq 0\right\}\;, 
\end{eqnarray}
for $C_{1}D_{1}\neq0$
\begin{equation}\label{18a}
\frac{D_{2}}{(L-a-\epsilon/2) D_{1}}\leq 0\;,\quad C_{1}=0\;,\quad\textrm{and}\quad \frac{C_{2}}{(L-a-\epsilon/2) C_{1}}\leq 0\;,\quad D_{1}=0\;,
\end{equation}   
and for $C_{1}=D_{1}=0$.
Equation (\ref{12}), instead, has only real solutions if 
\begin{equation}\label{19}
\left\{\frac{B_{2}C_{2}}{\epsilon^{2}B_{1}C_{1}}\leq 0\;,\frac{1}{\epsilon}\left(\frac{B_{2}}{ B_{1}}+\frac{C_{2}}{C_{1}}\right)\leq -1\right\}\;,\quad \textrm{or}\quad
\left\{\frac{B_{2}}{\epsilon B_{1}}\geq 0\;, \frac{C_{2}}{\epsilon C_{1}}\geq 0\right\}\;, 
\end{equation}
when $B_{1}C_{1}\neq0$
\begin{equation}\label{19a}
\frac{B_{2}}{\epsilon B_{1}}\geq 0\;,\quad C_{1}=0\;,\quad\textrm{and}\quad \frac{C_{2}}{\epsilon C_{1}}\geq 0\;,\quad B_{1}=0\;,
\end{equation}
and for $C_{1}=B_{1}=0$.
Since, as already mentioned earlier, we are only interested in real eigenvalues $\alpha_{j}$ we assume for the rest of this work
that the coefficients of the boundary conditions in (\ref{5}), (\ref{6}), and (\ref{7}) satisfy the above conditions.

The equations in (\ref{14})-(\ref{16}) play a fundamentally important role in expressing the spectral zeta function in each region
in terms of a complex integral \cite{kirsten01}. In our case we have
\begin{equation}\label{20}
\zeta_{j}(s,\epsilon,a)=\sum_{\lambda}n(\lambda)\zeta_{j}^{\lambda}(s,\epsilon,a)\;,
\end{equation} 
where $n(\lambda)$ represents the multiplicity of the eigenvalues of the Laplacian $\Delta_{N}$, and the functions $\zeta_{j}^{\lambda}(s,\epsilon,a)$
are defined as
\begin{equation}\label{21}
\zeta_{j}^{\lambda}(s,\epsilon,a)=\frac{\sin\pi s}{\pi}\lambda^{-2s}\int_{0}^{\infty}z^{-2s}\frac{\partial}{\partial z}\ln f_{\lambda}^{j}(i\lambda z,\epsilon,a)\;,
\end{equation}
with $f_{\lambda}^{j}(iz,\epsilon,a)$ given in (\ref{14}), (\ref{15}), and (\ref{16}) for $j=I$, $j=II$, and $j=III$, respectively.
It can be proved, by analyzing the asymptoic behavior of $f_{\lambda}^{j}(i\lambda z,\epsilon,a)$ as $z\to 0$ and as $z\to\infty$, that
the integral representation (\ref{21}) is valid in the region of the complex plane $1/2<\Re(s)<1$ \cite{fucci15}. 

The Casimir energy, and hence the force, is obtained from the knowledge of the spectral zeta function in the neighborhood of
$s=-1/2$. Since this point does not belong to the region $1/2<\Re(s)<1$, the zeta function in (\ref{20}) needs to be 
analytically continued to the complex semi-plane $\Re(s)\leq 1/2$.

Before concluding this Section we would like to discuss the limit of infinitely thin pistons. 
According to the boundary conditions (\ref{5})-(\ref{7}) the left side and the right side of the 
thick piston are allowed to have different types of boundary conditions. However, in the limit of a thin piston, $\epsilon\to 0$,
it is reasonable to assume that the boundary conditions to the left and right side of the piston must coalesce. 
This implies that results regarding infinitely thin pistons cannot be obtained by simply taking the limit $\epsilon\to 0$ of the ones for thick pistons.
In fact the correct thin piston limit is obtained not only by performing the 
limit $\epsilon\to 0$ but also by imposing the following relations $B_{2}=-C_{2}$ and $B_{1}=C_{1}$ on the parameters describing the 
boundary conditions on the piston itself.

\section{Analytic Continuation and the Casimir Energy }\label{sec3}

The restriction $\Re(s)>1/2$ indicated in the previous Section is due to the behavior of the integrand in (\ref{21}) as $\lambda z\to\infty$. 
In order to extend the spectral zeta function (\ref{20}) to the region to the left of $\Re(s)=1/2$, one needs to subtract and add an
appropriate number of terms of the asymptotic expansion of $\ln f_{\lambda}^{j}(i\lambda z,\epsilon,a)$ as $\lambda\to\infty$ uniform in the 
variable $z$. To obtain the desired asymptotic expansion we use the exponential form of the hyperbolic functions to write $\ln f_{\lambda}^{j}(i\lambda z,\epsilon,a)$ 
in region $I$ as
\begin{eqnarray}\label{22}
\ln f^{I}_{\lambda}\left(i\lambda z,a-\frac{\epsilon}{2}\right)&=&\lambda\sqrt{1+z^{2}}\left(a-\frac{\epsilon}{2}\right)+\ln\left[\frac{A_{1}B_{1}}{2\lambda\sqrt{1+z^{2}}}-\frac{A_{1}B_{2}+A_{2}B_{1}}{2}+\frac{A_{2}B_{2}}{2}\lambda\sqrt{1+z^{2}}\right]\nonumber\\
&+&\ln\left[1+\varepsilon_{I}(\lambda z,\epsilon,a)\right]\;,
\end{eqnarray}  
in region $II$ as 
\begin{eqnarray}\label{23}
\ln f^{II}_{\lambda}\left(i\lambda z,\epsilon\right)&=&\lambda\sqrt{1+z^{2}}\epsilon+\ln\left[\frac{B_{1}C_{1}}{2\lambda\sqrt{1+z^{2}}}+\frac{B_{1}C_{2}+B_{2}C_{1}}{2}+\frac{B_{2}C_{2}}{2}\lambda\sqrt{1+z^{2}}\right]\nonumber\\
&+&\ln\left[1+\varepsilon_{II}(\lambda z,\epsilon,a)\right]\;,
\end{eqnarray}
and, finally, in region $III$ as
\begin{eqnarray}\label{24}
\ln f^{III}_{\lambda}\left(i\lambda z,L-a-\frac{\epsilon}{2}\right)&=&\lambda\sqrt{1+z^{2}}\left(L-a-\frac{\epsilon}{2}\right)+\ln\left[\frac{C_{1}D_{1}}{2\lambda\sqrt{1+z^{2}}}-\frac{C_{1}D_{2}+C_{2}D_{1}}{2}+\frac{C_{2}D_{2}}{2}\lambda\sqrt{1+z^{2}}\right]\nonumber\\
&+&\ln\left[1+\varepsilon_{III}(\lambda z,\epsilon,a)\right]\;,
\end{eqnarray} 
where the functions $\varepsilon_{j}(\lambda z,\epsilon,a)$ describe the exponentially small terms.
The specific form of the uniform asymptotic expansion depends on the value of some of the parameters in the boundary conditions (\ref{5})-(\ref{7}) \cite{fucci15a}.
For instance, the asymptotic expansion for $\ln f^{I}_{\lambda}(i\lambda z,a-\epsilon/2)$ as $\lambda\to\infty$ uniform in $z$ depends on whether the coefficients $A_{2}$,
$B_{2}$ or both are either zero or different than zero. The possible combinations lead to four different forms of the uniform asymptotic expansion of $\ln f^{I}_{\lambda}(i\lambda z,a-\epsilon/2)$. Obviously, the same remarks hold true for $\ln f^{II}_{\lambda}(i\lambda z,\epsilon)$ with the parameters $B_{2}$ and $C_{2}$, and for 
$\ln f^{III}_{\lambda}(i\lambda z,L-a-\epsilon/2)$ with the parameters $C_{2}$ and $D_{2}$. 

A detailed description on how to obtain the uniform asymptotic expansion of $\ln f_{\lambda}^{j}(i\lambda z,\epsilon,a)$ can be found in
\cite{fucci15} and, for the sake of brevity, will not be repeated here. We focus, instead, on the final results. By introducing the function
\begin{equation}
\delta(x)=\left\{\begin{array}{ll}
1 & \textrm{if}\; x=0\\
0 & \textrm{if}\; x\neq 0
\end{array}\right.\;,
\end{equation}
we can write the different forms of the uniform asymptotic expansion in one expression. In particular, one obtains, in region $I$,  
\begin{eqnarray}\label{25}
\ln f^{I}_{\lambda}\left(i\lambda z,a-\frac{\epsilon}{2}\right)&\sim&\lambda\sqrt{1+z^{2}}\left(a-\frac{\epsilon}{2}\right)+\left[1-\delta(A_{2}B_{2})-\delta(A_{2})\delta(B_{2})\right]\ln\left(\lambda\sqrt{1+z^{2}}\right)\nonumber\\
&+&\left[1-\delta(A_{2}B_{2})\right]\ln\left(\frac{A_{2}B_{2}}{2}\right)
+\delta(A_{2})\delta(B_{2})\ln\left(\frac{A_{1}B_{1}}{2}\right)\nonumber\\
&+&\left[\delta(A_{2}B_{2})-\delta(A_{2})\delta(B_{2})\right]\ln\left(-\frac{A_{1}B_{2}\delta(A_{2})+A_{2}B_{1}\delta(B_{2})}{2}\right)
+\sum_{k=1}^{\infty}\frac{g^{I}_{k}}{\lambda^{k}(1+z^{2})^{\frac{k}{2}}}\;,
\end{eqnarray}
where
\begin{equation}\label{26}
g^{I}_{k}=\left(\frac{\delta(A_{2}B_{2})-1}{k}\right)\left[\left(\frac{A_{1}}{A_{2}}\right)^{k}+\left(\frac{B_{1}}{B_{2}}\right)^{k}\right]
+\left(\frac{\delta(A_{2}B_{2})-\delta(A_{2})\delta(B_{2})}{k}\right)\left(\frac{A_{1}B_{1}}{A_{1}B_{2}\delta(A_{2})+A_{2}B_{1}\delta(B_{2})}\right)^{k}\;,
\end{equation}
in region $II$,
\begin{eqnarray}\label{27}
\ln f^{II}_{\lambda}\left(i\lambda z,\epsilon\right)&\sim&\lambda\sqrt{1+z^{2}}\epsilon+\left[1-\delta(B_{2}C_{2})-\delta(B_{2})\delta(C_{2})\right]\ln\left(\lambda\sqrt{1+z^{2}}\right)
+\left[1-\delta(B_{2}C_{2})\right]\ln\left(\frac{B_{2}C_{2}}{2}\right)\nonumber\\
&+&\delta(B_{2})\delta(C_{2})\ln\left(\frac{B_{1}C_{1}}{2}\right)+\left[\delta(B_{2}C_{2})-\delta(B_{2})\delta(C_{2})\right]\ln\left(\frac{B_{1}C_{2}\delta(B_{2})+B_{2}C_{1}\delta(C_{2})}{2}\right)\nonumber\\
&+&\sum_{k=1}^{\infty}\frac{(-1)^{k}g^{II}_{k}}{\lambda^{k}(1+z^{2})^{\frac{k}{2}}}\;,
\end{eqnarray}
with
\begin{equation}\label{28}
g^{II}_{k}=\left(\frac{\delta(B_{2}C_{2})-1}{k}\right)\left[\left(\frac{B_{1}}{B_{2}}\right)^{k}+\left(\frac{C_{1}}{C_{2}}\right)^{k}\right]
+\left(\frac{\delta(B_{2}C_{2})-\delta(B_{2})\delta(C_{2})}{k}\right)\left(\frac{B_{1}C_{1}}{B_{1}C_{2}\delta(B_{2})+B_{2}C_{1}\delta(C_{2})}\right)^{k}\;,
\end{equation}
and in region $III$
\begin{eqnarray}\label{29}
\ln f^{III}_{\lambda}\left(i\lambda z,L-a-\frac{\epsilon}{2}\right)&\sim&\lambda\sqrt{1+z^{2}}\left(L-a-\frac{\epsilon}{2}\right)+\left[1-\delta(C_{2}D_{2})-\delta(C_{2})\delta(D_{2})\right]\ln\left(\lambda\sqrt{1+z^{2}}\right)\nonumber\\
&+&\left[1-\delta(C_{2}D_{2})\right]\ln\left(\frac{C_{2}D_{2}}{2}\right)
+\delta(C_{2})\delta(D_{2})\ln\left(\frac{C_{1}D_{1}}{2}\right)\nonumber\\
&+&\left[\delta(C_{2}D_{2})-\delta(C_{2})\delta(D_{2})\right]\ln\left(-\frac{C_{1}D_{2}\delta(C_{2})+C_{2}D_{1}\delta(D_{2})}{2}\right)\nonumber\\
&+&\sum_{k=1}^{\infty}\frac{g^{III}_{k}}{\lambda^{k}(1+z^{2})^{\frac{k}{2}}}\;,
\end{eqnarray}
where
\begin{equation}\label{30}
g^{III}_{k}=\left(\frac{\delta(C_{2}D_{2})-1}{k}\right)\left[\left(\frac{C_{1}}{C_{2}}\right)^{k}+\left(\frac{D_{1}}{D_{2}}\right)^{k}\right]
+\left(\frac{\delta(C_{2}D_{2})-\delta(C_{2})\delta(D_{2})}{k}\right)\left(\frac{C_{1}D_{1}}{C_{1}D_{2}\delta(C_{2})+C_{2}D_{1}\delta(D_{2})}\right)^{k}\;.
\end{equation}
 
The explicit knowledge of the uniform asymptotic expansions (\ref{25}), (\ref{27}), and (\ref{29}) allows us to proceed with the analytic
continuation of the spectral zeta function $\zeta_{j}(s,\epsilon,a)$ in (\ref{20}). By subtracting and adding, in the integral (\ref{21}) 
representing $\zeta^{\lambda}_{j}(s,\epsilon,a)$,  $N$ terms of the expansion for $\ln f_{\lambda}^{j}(i\lambda z,\epsilon,a)$ we obtain 
the analytically continued expression for the spectral zeta function in (\ref{20}) of the form
\begin{equation}\label{31}
\zeta_{j}(s,\epsilon,a)= Z_{j}(s,\epsilon,a)+\sum_{i=-1}^{N} R^{j}_{i}(s,\epsilon,a)\;.
\end{equation} 
By construction, the function $Z_{j}(s,\epsilon,a)$ is analytic in the semi-plane $\Re(s)>(d-N-1)/2$ and $R^{j}_{i}(s,\epsilon,a)$ is a meromorphic 
function of $s\in\mathbb{C}$. Clearly, their specific expression depends on the region $j$. By introducing the spectral zeta function 
associated with the Laplacian on the manifold $N$
\begin{equation}\label{32}
\zeta_{N}(s)=\sum_{\lambda}n(\lambda)\lambda^{-2s}\;,
\end{equation} 
in region $I$ we have
\begin{eqnarray}\label{33}
 Z_{I}(s,\epsilon,a)&=&\frac{\sin\pi s}{\pi}\sum_{\lambda}n(\lambda)\lambda^{-2s}\int_{0}^{\infty}z^{-2s}\Bigg[\frac{\partial}{\partial z}\ln f^{I}_{\lambda}\left(i\lambda z,a-\frac{\epsilon}{2}\right)
-\lambda\sqrt{1+z^{2}}\left(a-\frac{\epsilon}{2}\right)\nonumber\\
&-&\left[1-\delta(A_{2}B_{2})\right]\ln\left(\frac{A_{2}B_{2}}{2}\right)
-\left[1-\delta(A_{2}B_{2})-\delta(A_{2})\delta(B_{2})\right]\ln\left(\lambda\sqrt{1+z^{2}}\right)\nonumber\\
&-&\delta(A_{2})\delta(B_{2})\ln\left(\frac{A_{1}B_{1}}{2}\right)
-\left[\delta(A_{2}B_{2})-\delta(A_{2})\delta(B_{2})\right]\ln\left(-\frac{A_{1}B_{2}\delta(A_{2})+A_{2}B_{1}\delta(B_{2})}{2}\right)\nonumber\\
&-&\sum_{k=1}^{N}\frac{g^{I}_{k}}{\lambda^{k}(1+z^{2})^{\frac{k}{2}}}\Bigg]\diff z\;,
\end{eqnarray}
with
\begin{equation}\label{34a}
R^{I}_{-1}(s,\epsilon,a)=\frac{a-\epsilon/2}{2\sqrt{\pi}\Gamma(s)}\Gamma\left(s-\frac{1}{2}\right)\zeta_{N}\left(s-\frac{1}{2}\right)\;,\quad 
R^{I}_{0}(s,\epsilon,a)=\frac{1}{2}\left[1-\delta(A_{2}B_{2})-\delta(A_{2})\delta(B_{2})\right]\zeta_{N}(s)\;,
\end{equation}
and, for $i\geq 1$,
\begin{equation}\label{35a}
R^{I}_{i}(s,\epsilon,a)=-\frac{g^{I}_i}{\Gamma\left(\frac{i}{2}\right)\Gamma(s)}\Gamma\left(s+\frac{i}{2}\right)\zeta_{N}\left(s+\frac{i}{2}\right)\;.
\end{equation}
In region $II$ we find, instead,
\begin{eqnarray}\label{36a}
 Z_{II}(s,\epsilon)&=&\frac{\sin\pi s}{\pi}\sum_{\lambda}n(\lambda)\lambda^{-2s}\int_{0}^{\infty}z^{-2s}\Bigg[\frac{\partial}{\partial z}\ln f^{II}_{\lambda}\left(i\lambda z,\epsilon\right)
-\lambda\sqrt{1+z^{2}}\epsilon-\left[1-\delta(B_{2}C_{2})\right]\ln\left(\frac{B_{2}C_{2}}{2}\right)\nonumber\\
&-&\left[1-\delta(B_{2}C_{2})-\delta(B_{2})\delta(C_{2})\right]\ln\left(\lambda\sqrt{1+z^{2}}\right)
-\delta(B_{2})\delta(C_{2})\ln\left(\frac{B_{1}C_{1}}{2}\right)\nonumber\\
&-&\left[\delta(B_{2}C_{2})-\delta(B_{2})\delta(C_{2})\right]\ln\left(\frac{B_{1}C_{2}\delta(B_{2})+B_{2}C_{1}\delta(C_{2})}{2}\right)
-\sum_{k=1}^{N}\frac{(-1)^{k}g^{II}_{k}}{\lambda^{k}(1+z^{2})^{\frac{k}{2}}}\Bigg]\diff z\;,
\end{eqnarray}
where
\begin{equation}\label{37b}
R^{II}_{-1}(s,\epsilon,a)=\frac{\epsilon}{2\sqrt{\pi}\Gamma(s)}\Gamma\left(s-\frac{1}{2}\right)\zeta_{N}\left(s-\frac{1}{2}\right)\;,\quad
R^{II}_{0}(s,\epsilon,a)=\frac{1}{2}\left[1-\delta(B_{2}C_{2})-\delta(B_{2})\delta(C_{2})\right]\zeta_{N}(s)\;,
\end{equation}
and
\begin{equation}\label{38a}
R^{II}_{i}(s,\epsilon,a)=\frac{(-1)^{i+1}g^{II}_i}{\Gamma\left(\frac{i}{2}\right)\Gamma(s)}\Gamma\left(s+\frac{i}{2}\right)\zeta_{N}\left(s+\frac{i}{2}\right)\;,
\end{equation}
for for $i\geq 1$.
Finally, in region $III$ we get
\begin{eqnarray}\label{39a}
 Z_{III}(s,\epsilon,a)&=&\frac{\sin\pi s}{\pi}\sum_{\lambda}n(\lambda)\lambda^{-2s}\int_{0}^{\infty}z^{-2s}\Bigg[\frac{\partial}{\partial z}\ln f^{III}_{\lambda}\left(i\lambda z,L-a-\frac{\epsilon}{2}\right)
-\lambda\sqrt{1+z^{2}}\left(L-a-\frac{\epsilon}{2}\right)\nonumber\\
&-&\left[1-\delta(C_{2}D_{2})\right]\ln\left(\frac{C_{2}D_{2}}{2}\right)
-\left[1-\delta(C_{2}D_{2})-\delta(C_{2})\delta(D_{2})\right]\ln\left(\lambda\sqrt{1+z^{2}}\right)\nonumber\\
&-&\delta(C_{2})\delta(D_{2})\ln\left(\frac{C_{1}D_{1}}{2}\right)
-\left[\delta(C_{2}D_{2})-\delta(C_{2})\delta(D_{2})\right]\ln\left(-\frac{C_{1}D_{2}\delta(C_{2})+C_{2}D_{1}\delta(D_{2})}{2}\right)\nonumber\\
&-&\sum_{k=1}^{N}\frac{g^{III}_{k}}{\lambda^{k}(1+z^{2})^{\frac{k}{2}}}\Bigg]\diff z\;,
\end{eqnarray}
with
\begin{equation}\label{34}
R^{III}_{-1}(s,\epsilon,a)=\frac{L-a-\epsilon/2}{2\sqrt{\pi}\Gamma(s)}\Gamma\left(s-\frac{1}{2}\right)\zeta_{N}\left(s-\frac{1}{2}\right)\;,\quad 
R^{III}_{0}(s,\epsilon,a)=\frac{1}{2}\left[1-\delta(C_{2}D_{2})-\delta(C_{2})\delta(D_{2})\right]\zeta_{N}(s)\;,
\end{equation}
and, for $i\geq 1$,
\begin{equation}\label{35}
R^{III}_{i}(s,\epsilon,a)=-\frac{g^{III}_i}{\Gamma\left(\frac{i}{2}\right)\Gamma(s)}\Gamma\left(s+\frac{i}{2}\right)\zeta_{N}\left(s+\frac{i}{2}\right)\;.
\end{equation}
It is important to make a remark at this point. In performing the analytic continuation of the spectral zeta function 
we have assumed, without explicitly mentioning it, that the Laplacian on manifold $N$ possesses no zero modes. 
If a zero mode is present, then the process of analytic continuation has to be modified. This is due to the fact that the 
asymptotic expansions utilized above become somewhat different when dealing with zero modes. The interested reader can find the details of this case
reported in \cite{fucci15}.   

By setting $N=D$ in (\ref{31}) and by using (\ref{3}) we obtain an expression for the spectral zeta function
associated with the thick piston valid in the region $-1<\Re(s)<1$ and, hence, suitable for the analysis of the Casimir energy. 
In more details we have
\begin{eqnarray}\label{36}
\zeta(s,\epsilon,a)&=&Z_{I}(s,\epsilon,a)+Z_{II}(s,\epsilon)+Z_{III}(s,\epsilon,a)+\frac{L}{2\sqrt{\pi}\Gamma(s)}\Gamma\left(s-\frac{1}{2}\right)\zeta_{N}\left(s-\frac{1}{2}\right)\nonumber\\
&+&\frac{1}{2}\left[3-\delta(A_{2}B_{2})-\delta(B_{2}C_{2})-\delta(C_{2}D_{2})-\delta(A_{2})\delta(B_{2})-\delta(B_{2})\delta(C_{2})-\delta(C_{2})\delta(D_{2})\right]\zeta_{N}(s)\nonumber\\
&-&\sum_{i=1}^{D}\frac{g^{I}_{i}+(-1)^{i}g^{II}_{i}+g^{III}_{i}}{\Gamma\left(\frac{i}{2}\right)\Gamma(s)}\Gamma\left(s+\frac{i}{2}\right)\zeta_{N}\left(s+\frac{i}{2}\right)\;.
\end{eqnarray}
The Casimir energy is obtained from this expression by first setting $s=\omega-1/2$ and by then performing the asymptotic expansion as $\omega\to 0$. 
To complete this task special attention needs to be paid to the meromorphic structure of $\zeta_{N}(s)$ which, according to the 
theory of spectral zeta functions \cite{gilkey95,kirsten01}, reads
\begin{eqnarray}\label{37}
\zeta_{N}(\omega -n)&=&\frac{(-1)^{n}n!}{(4\pi)^{\frac{d}{2}}}A^{N}_{\frac{d}{2}+n}+\omega\zeta^{\prime}_{N}(-n)+O(\omega^{2})\;,\\
\zeta_{N}\left(\omega+\frac{d-k}{2}\right)&=&\frac{A^{N}_{k/2}}{\omega\Gamma\left(\frac{d-k}{2}\right)}+\textrm{FP}\,\zeta_{N}\left(\frac{d-k}{2}\right)+O(\omega)\;,\\
\zeta_{N}\left(\omega-\frac{2n+1}{2}\right)&=&\frac{A^{N}_{(d+2n+1)/2}}{\omega\Gamma\left(-\frac{2n+1}{2}\right)}+\textrm{FP}\,\zeta_{N}\left(-\frac{2n+1}{2}\right)+O(\omega)\;,\label{37a}
\end{eqnarray}
where $n\in\mathbb{N}_{0}$, $k=\{0,\ldots,d-1\}$, and $A^{N}_{m/2}$ are the coefficients of the small-$t$ asymptotic expansion
of the trace of the heat kernel associated with the Laplacian $\Delta_{N}$ \cite{gilkey95,vassilevich03}.

By using the formula (\ref{36}) and the pole structure displayed in (\ref{37})-(\ref{37a}) we obtain for the Casimir energy of
the thick piston the following expression
\begin{eqnarray}\label{38}
\lefteqn{E_{\textrm{Cas}}(\epsilon,a)=\frac{1}{2}\left(\frac{1}{\omega}+\ln\mu^{2}\right)\Bigg\{-\frac{1}{4\sqrt{\pi}}[3-\delta(A_{2}B_{2})-\delta(B_{2}C_{2})-\delta(C_{2}D_{2})-\delta(A_{2})\delta(B_{2})-\delta(B_{2})\delta(C_{2})}\nonumber\\
&-&\delta(C_{2})\delta(D_{2})]A^{N}_{\frac{d+1}{2}}-\frac{L}{(4\pi)^{\frac{d+2}{2}}}A^{N}_{\frac{d+2}{2}}+2\frac{g^{I}_{1}-g^{II}_{1}+g^{III}_{1}}{(4\pi)^{\frac{d+2}{2}}}A^{N}_{\frac{d}{2}}+\sum_{k=2}^{D}\frac{g^{I}_{k}+(-1)^{k}g^{II}_{k}+g^{III}_{k}}{2\sqrt{\pi}\Gamma\left(\frac{k}{2}\right)}A^{N}_{\frac{d+1-k}{2}}\Bigg\}\nonumber\\
&+&\frac{1}{2} Z_{I}\left(-\frac{1}{2},\epsilon,a\right)+\frac{1}{2}Z_{II}\left(-\frac{1}{2},\epsilon\right)+\frac{1}{2}Z_{III}\left(-\frac{1}{2},\epsilon,a\right)+\frac{L}{8\pi}\left[\zeta'_{N}(-1)-\frac{2\ln 2-1}{(4\pi)^{\frac{d}{2}}}A^{N}_{\frac{d+2}{2}}\right]\nonumber\\
&+&\frac{1}{4}\left[3-\delta(A_{2}B_{2})-\delta(B_{2}C_{2})-\delta(C_{2}D_{2})-\delta(A_{2})\delta(B_{2})-\delta(B_{2})\delta(C_{2})-\delta(C_{2})\delta(D_{2})\right]\textrm{FP}\,\zeta_{N}\left(-\frac{1}{2}\right)\nonumber\\
&+&\frac{g^{I}_{1}-g^{II}_{1}+g^{III}_{1}}{4\pi}\left[\zeta'_{N}(0)+\frac{2(\ln 2-1)}{(4\pi)^{\frac{d}{2}}}A^{N}_{\frac{d}{2}}\right]\nonumber\\
&+&\sum_{k=2}^{D}\frac{g^{I}_{k}+(-1)^{k}g^{II}_{k}+g^{III}_{k}}{4\sqrt{\pi}\Gamma\left(\frac{k}{2}\right)}\left[\Gamma\left(\frac{k-1}{2}\right)\textrm{FP}\,\zeta_{N}\left(\frac{k-1}{2}\right)+
\left(2-\gamma-2\ln 2+\Psi\left(\frac{k-1}{2}\right)\right)A^{N}_{\frac{d+1-k}{2}}\right]+O(\omega)\;.\nonumber\\
\end{eqnarray}
It is well-known that the Casimir energy of a piston is generally not a well defined quantity.
This feature can clearly be seen from the above expression: In fact, the ambiguity in the Casimir energy is 
explicitly dependent on the heat kernel coefficients $A^{N}_{(d+1-k)/2}$ with $k=\{-2,\ldots,d+1\}$ and, consequently,
on the geometry of the manifold $N$.

\section{The Casimir force on particular thick pistons}\label{sec4}

Although for a general piston configuration the Casimir energy is not well defined, the Casimir 
force on the piston itself can be proved to be unambiguous. 
From the definition of the Casimir force in (\ref{0b}), it is not very difficult to obtain, from (\ref{38}),
\begin{equation}\label{39}
F_{\textrm{Cas}}(\epsilon,a)=-\frac{1}{2} Z'_{I}\left(-\frac{1}{2},\epsilon,a\right)-\frac{1}{2}Z'_{III}\left(-\frac{1}{2},\epsilon,a\right)\;,
\end{equation}
where the prime denotes differentiation with respect to the position of the piston $a$. We would like to point out that since
the function $Z_{II}$ is independent of the parameter $a$, there is no contribution, as one would expect, to the Casimir force on the 
thick piston coming from region $II$ (namely the piston itself). 
The formulas obtained in (\ref{38}) for the 
Casimir energy and in (\ref{39}) for the corresponding force on the piston are valid for any smooth, compact Riemannian manifold $N$ and for separated boundary conditions
satisfying the constraints (\ref{17}) through (\ref{19a}). In order to study more explicit examples of Casimir force on the thick piston 
we need to specify the manifold $N$. In this Section we assume that the cross-section $N$ of the piston configuration is a $d$-dimensional sphere. 
In this circumstance the eigenvalues $\lambda$ associated with the Laplacian on $N$ are known to be
\begin{equation}\label{40}
\lambda=l-\frac{d-1}{2}\;,
\end{equation}  
with $\in\mathbb{N}_{0}$. It can also be shown that each eigenvalue in (\ref{40}) has multiplicity 
\begin{equation}\label{41}
n(l)=(2l+d-1)\frac{(l+d-2)!}{l!(d-1)!}\;.
\end{equation}  

We can now use the specific formulas (\ref{40}) and (\ref{41}) in (\ref{39}) and analyze how the Casimir force on a thick piston 
with spherical cross-section behaves as the boundary conditions change within the constraints (\ref{17})-(\ref{19a}). Due to the form of 
$Z_{j}\left(s,\epsilon,a\right)$ it is clear that that the Casimir force has to be studied numerically.
At this point, however, it is important to make a remark. From the expressions (\ref{33}), (\ref{36}), and (\ref{39}) it is apparent that functions $Z_{I}$, $Z_{II}$, and
$Z_{III}$, and consequantly the Casimir force in (\ref{39}), depend explicitly not only on the position of the piston $a$ and its thickness $\epsilon$, but also on the parameters 
describing the boundary conditions in (\ref{5}), (\ref{6}), and (\ref{7}).
Any attempt at a numerical study of the behavior of the Casimir force on a thick piston as all the parameters vary simultaneously 
would lead to quite complex results that would not show the role that each parameter plays in the Casimir force. 
For this reason, in order simplify the analysis, we fix some of the parameters and let the remaining ones be free to vary according 
to the constraints (\ref{17})-(\ref{19a}). In what follows we will work under the assumption that the manifold $N$ is of dimension $d=2$,
and without loss of generality, we will consider a piston configuration of unit length $L=1$.

The graphs we provide next in this Section display the Casimir force on the thick piston as 
a function of the position $a$ in a number of particular cases. 
The lines of different thickness in each graph provide a plot of the Casimir force on pistons of varying thickness: thicker lines 
represent the force on thicker pistons. In all the examples outlined in this Section, we have considered the following values: $\epsilon=0.01$, 
$\epsilon=0.05$, $\epsilon=0.2$, and $\epsilon=0.5$.
For each example considered we also examine the limit of infinitely thin pistons $\epsilon\to 0$. 
The infinitely thin piston case has been analyzed in details in \cite{fucci15} and the results found there
have been used to obtain the plots, displayed with dashed lines, of the Casimir force as a function of the position of the piston $a$.

\subsection{Dirichlet Boundary Conditions}

\begin{figure}[]
\centering
\includegraphics[scale=0.85,trim=0cm 0cm 0cm 0cm, clip=true, angle=0]{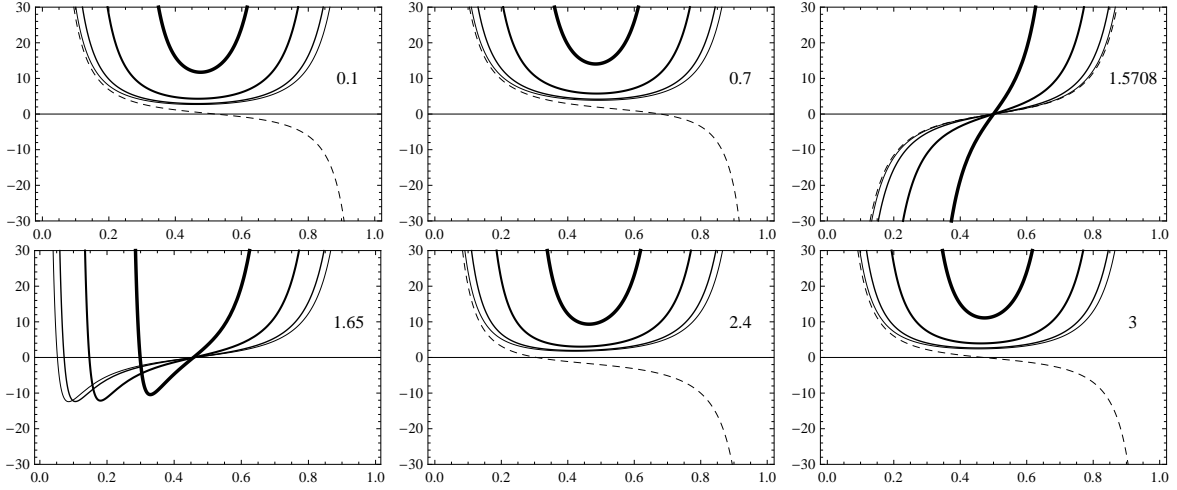}
\caption{Dirichlet boundary conditions at the endpoints: DGDD case. Each graph is obtained by fixing the value of the parameter $\alpha$, displayed in the upper right corner, in the intervals $[0,\pi/4]$ and $[\pi/2,\pi]$. 
The parameter $a$ varies along the $x$-axis, while the Casimir force on the piston (in units for which $h=c=1$) varies along the $y$-axis.}\label{fig1}
\end{figure}

As a first example we consider the case in which Dirichlet boundary conditions are imposed on the left-end of the piston configuration, that is at $x=0$. 
On the left edge of the thick piston we impose general boundary conditions described by a parameter $\alpha$ while on its 
right edge and on the right-end of the piston configuration we have Dirichlet boundary conditions. This set-up, which we will refer to as DGDD, is expressed in terms of the parameters 
in the boundary conditions (\ref{5})-(\ref{7}) as follows
\begin{equation}\label{42}
A_{1}=1\;,A_{2}=0\;, B_{1}=\sin\alpha\;, B_{2}=\cos\alpha\;, C_{1}=1\;, C_{2}=0\;, D_{1}=1\;, D_{2}=0\;.
\end{equation}  
In this, as well as in the ensuing examples, we assume, without loss of generality, that $\alpha\in[0,\pi]$. In region $I$ the conditions (\ref{17}) provide the following
relations for the parameter $\alpha$
\begin{equation}\label{43}
\frac{\cot\alpha}{a-\epsilon/2}\geq 1\;, \quad \textrm{or}\quad \cot\alpha\leq 0\;.
\end{equation} 
These inequalities are satisfied uniformly in $a$ and $\epsilon$ if $\alpha\in[0,\pi/4]$ or $\alpha\in[\pi/2,\pi]$. The Casimir force on a thick piston for this configuration 
is given in Figure \ref{fig1}.
For infinitely thin pistons the parameter $\alpha$ can take values in the intervals $[0,\pi/4]$, $[3\pi/4,\pi]$ and take the value $\alpha=\pi/2$ \cite{fucci15}.  
This remark implies, in particular, that while the Casimir force on thick pistons is well defined for $\alpha\in(\pi/4,\pi/2)$ and $\alpha\in(\pi/2,3\pi/4)$,
the corresponding force on infinitely thin pistons is not. This is the reason why the dashed line is not present in one of the graphs in Figure \ref{fig1}.  

\begin{figure}[]
\centering
\includegraphics[scale=0.85,trim=0cm 0cm 0cm 0cm, clip=true, angle=0]{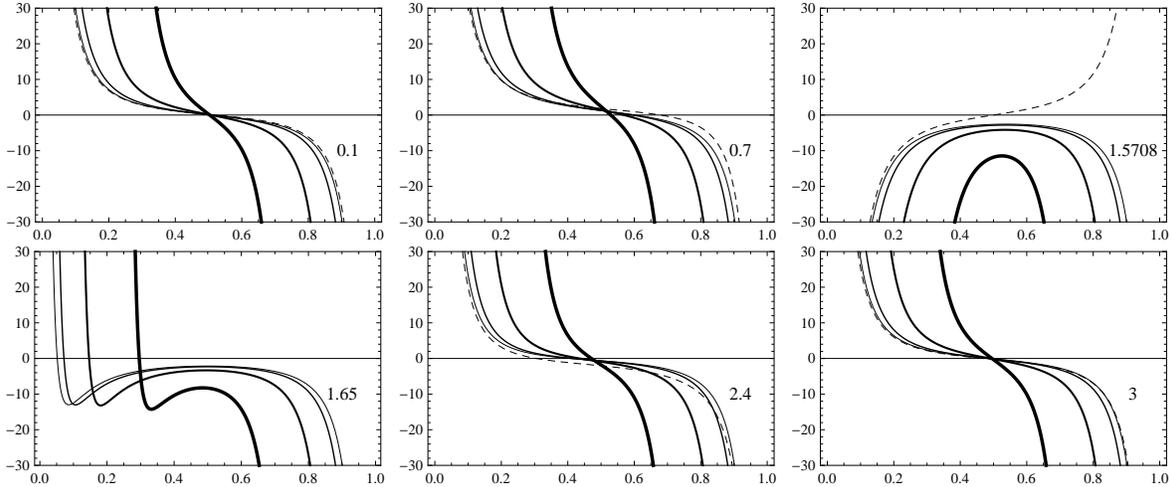}
\caption{Dirichlet boundary conditions at the endpoints: DGND case. Each graph is obtained by fixing the value of the parameter $\alpha$, displayed in the lower right corner, in the intervals $[0,\pi/4]$ and $[\pi/2,\pi]$. 
The parameter $a$ varies along the $x$-axis, while the Casimir force on the piston (in units for which $h=c=1$) varies along the $y$-axis. }\label{fig2}
\end{figure}

In the next example, we consider Dirichlet boundary conditions at both endpoints of the piston configuration $x=0$ and $x=1$. 
On the left edge of the thick piston we impose general boundary conditions, parametrized by $\alpha$, and on the right of the piston we consider 
Neumann boundary conditions. For brevity, we refer to this configuration as DGND and it is described, in terms of the parameters 
in the boundary conditions (\ref{5})-(\ref{7}), as
\begin{equation}\label{44}
A_{1}=1\;,A_{2}=0\;, B_{1}=\sin\alpha\;, B_{2}=\cos\alpha\;, C_{1}=0\;, C_{2}=1\;, D_{1}=1\;, D_{2}=0\;.
\end{equation}
According to the constraints (\ref{17}) the parameter $\alpha$ has to satisfy the inequalities (\ref{43}) which hold uniformly in $a$ and $\epsilon$ 
if, once again, $\alpha\in[0,\pi/4]$ or $\alpha\in[\pi/2,\pi]$. The Casimir force on a tick piston for this case is displayed in Figure \ref{fig2}.
The Casimir force on the corresponding infinitely thin piston, represented by the dashed lines in Figure \ref{fig2}, is the same as the one obtained in the DGDD case.

\subsection{Neumann Boundary Conditions}
 
\begin{figure}[]
\centering
\includegraphics[scale=0.85,trim=0cm 0cm 0cm 0cm, clip=true, angle=0]{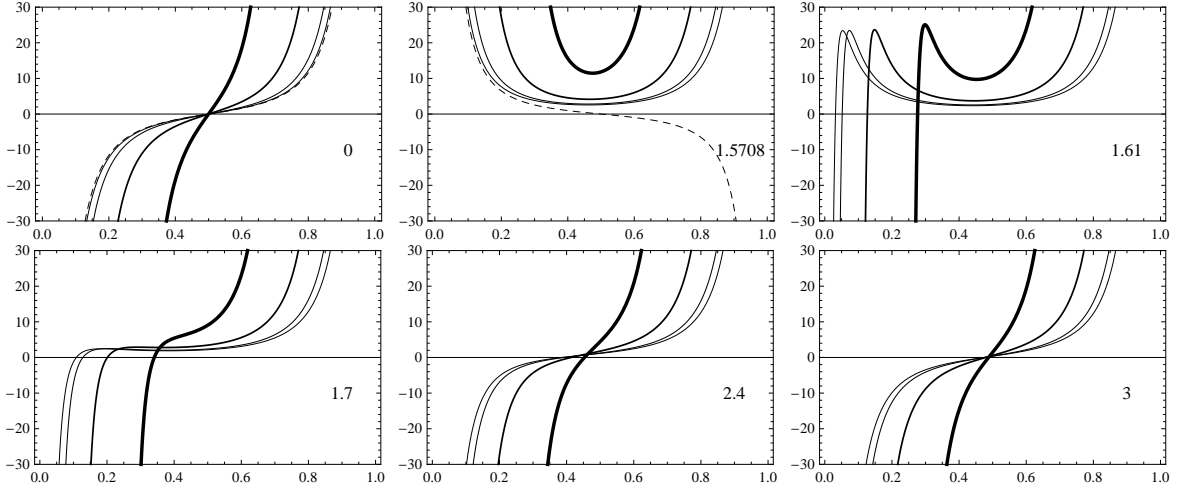}
\caption{Neumann boundary conditions at both endpoints: NGNN case. Each graph is obtained by fixing the value of the parameter $\alpha$, displayed in the lower right corner, in the interval $[\pi/2,\pi]$ and for $\alpha=0$. 
The parameter $a$ varies along the $x$-axis, while the Casimir force on the piston (in units for which $h=c=1$) varies along the $y$-axis. }\label{fig3}
\end{figure}

Here we analyze the case of Neumann boundary conditions at both endpoints of the piston configuration, namely at $x=0$ and $x=1$. On the left side of the thick piston 
we have general boundary conditions, described by the parameter $\alpha$, and on the right side of the piston we impose Neumann boundary condition. 
This configuration will be referred to as NGNN and is obtained from the conditions in (\ref{5})-(\ref{7}) by setting
\begin{equation}\label{45}
A_{1}=0\;,A_{2}=1\;, B_{1}=\sin\alpha\;, B_{2}=\cos\alpha\;, C_{1}=0\;, C_{2}=1\;, D_{1}=0\;, D_{2}=1\;.
\end{equation} 
The relations found in (\ref{17a}) lead to the following inequality
\begin{equation}\label{46}
\frac{\cot\alpha}{a-\epsilon/2}\leq 0\;,
\end{equation}  
which is satisfied for values of $\alpha$ in the interval $\alpha\in[\pi/2,\pi]$, and for $\alpha=0$. The Casimir 
force on a thick piston for this particular case is shown in Figure \ref{fig3}. The Casimir force on the corresponding
infinitely thin piston is well defined for the values $\alpha=0$ and $\alpha=\pi/2$ \cite{fucci15} 
and is displayed via dashed lines in 
Figure \ref{fig3}.

\begin{figure}[]
\centering
\includegraphics[scale=0.85,trim=0cm 0cm 0cm 0cm, clip=true, angle=0]{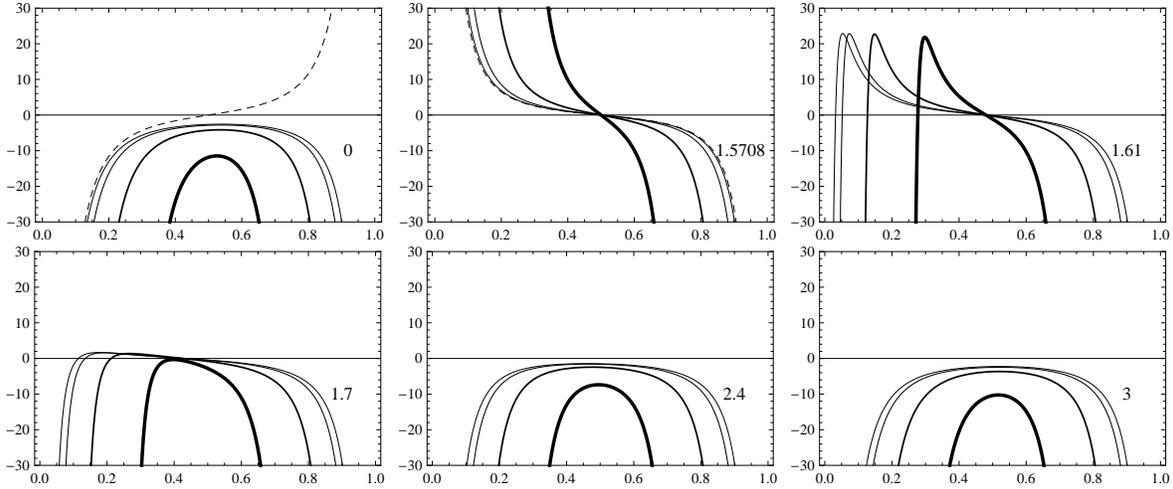}
\caption{Neumann boundary conditions at both endpoints: NGDN case. Each graph is obtained by fixing the value of the parameter $\alpha$, displayed in the lower right corner, in the interval $[\pi/2,\pi]$ and for $\alpha=0$. 
The parameter $a$ varies along the $x$-axis, while the Casimir force on the piston (in units for which $h=c=1$) varies along the $y$-axis.}\label{fig4}
\end{figure}

As a further example, we consider Neumann boundary conditions at the endpoints $x=0$ and $x=1$, general boundary conditions 
on the left side of the thick piston and Dirichlet boundary conditions on its right side. This configuration, which we denote by NGDN, is obtained by using the 
following values for the parameters in the boundary conditions (\ref{5})-(\ref{7})
\begin{equation}\label{47}
A_{1}=0\;,A_{2}=1\;, B_{1}=\sin\alpha\;, B_{2}=\cos\alpha\;, C_{1}=1\;, C_{2}=0\;, D_{1}=0\;, D_{2}=1\;.
\end{equation} 
Also in this example the parameter $\alpha$ must satisfy the inequality (\ref{46}) which implies that $\alpha\in[\pi/2,\pi]$ and $\alpha=0$. The Casimir force on a thick piston 
for this case is given in Figure \ref{fig4}.
Furthermore, we point out that the Casimir force on the corresponding infinitely thin piston, illustrated by the dashed lines in Figure \ref{fig3}, is the same 
as the one obtained in the NGNN case.

\subsection{Hybrid Boundary Conditions}

\begin{figure}[]
\centering
\includegraphics[scale=0.85,trim=0cm 0cm 0cm 0cm, clip=true, angle=0]{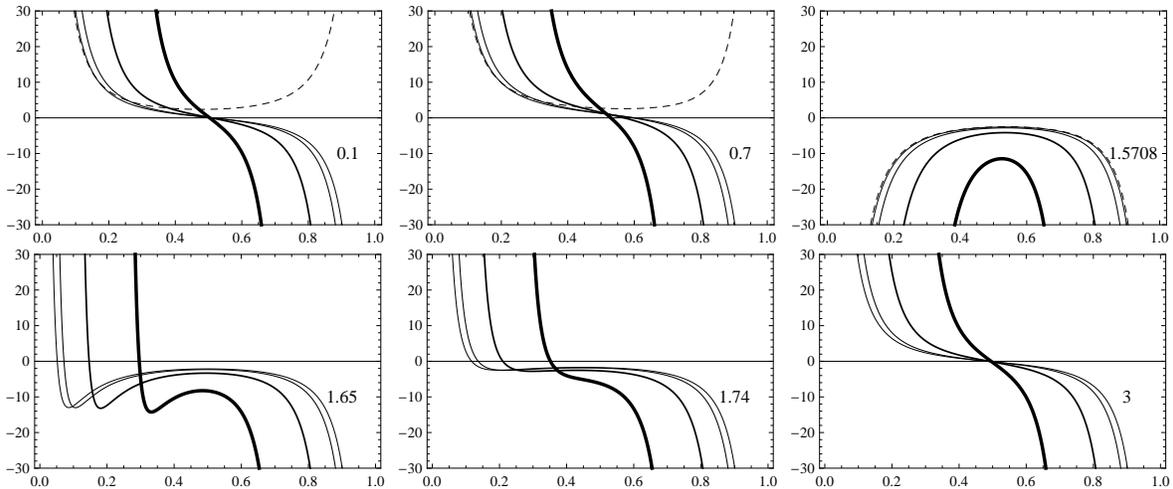}
\caption{Hybrid boundary conditions: DGDN case. Each graph is obtained by fixing the value of the parameter $\alpha$, 
displayed in the lower right corner, in the intervals $[0,\pi/4]$ and $[\pi/2,\pi]$. The parameter $a$ varies along the $x$-axis, 
while the Casimir force on the piston (in units for which $h=c=1$) varies along the $y$-axis. }\label{fig5}
\end{figure}

We consider now mixed boundary conditions. On the left end of the piston configuration we impose Dirichlet boundary conditions while on the right end we impose 
Neuman boundary conditions. On the left side of the thick piston we have general boundary conditions, parametrized by $\alpha$, and on the right side of the piston 
we have Dirichlet boundary conditions. This particular example will be denoted by DGDN and is obtained by setting in (\ref{5})-(\ref{7})
\begin{equation}\label{48}
A_{1}=1\;,A_{2}=0\;, B_{1}=\sin\alpha\;, B_{2}=\cos\alpha\;, C_{1}=1\;, C_{2}=0\;, D_{1}=0\;, D_{2}=1\;.
\end{equation}  
In this case the parameter $\alpha$ has to satisfy the inequality (\ref{43}) which occurs when $\alpha\in[0,\pi/4]$ or $\alpha\in[\pi/2,\pi]$. Graphs of the Casimir force 
on a thick piston for this example are provided in Figure \ref{fig5}.
The corresponding Casimir force on the infinitely thin piston is well defined for $\alpha\in[0,\pi/4]$ and for $\alpha=\pi/2$ \cite{fucci15}, and, once again,
is displayed in Figure \ref{fig5} using dashed lines.

\begin{figure}[]
\centering
\includegraphics[scale=0.85,trim=0cm 0cm 0cm 0cm, clip=true, angle=0]{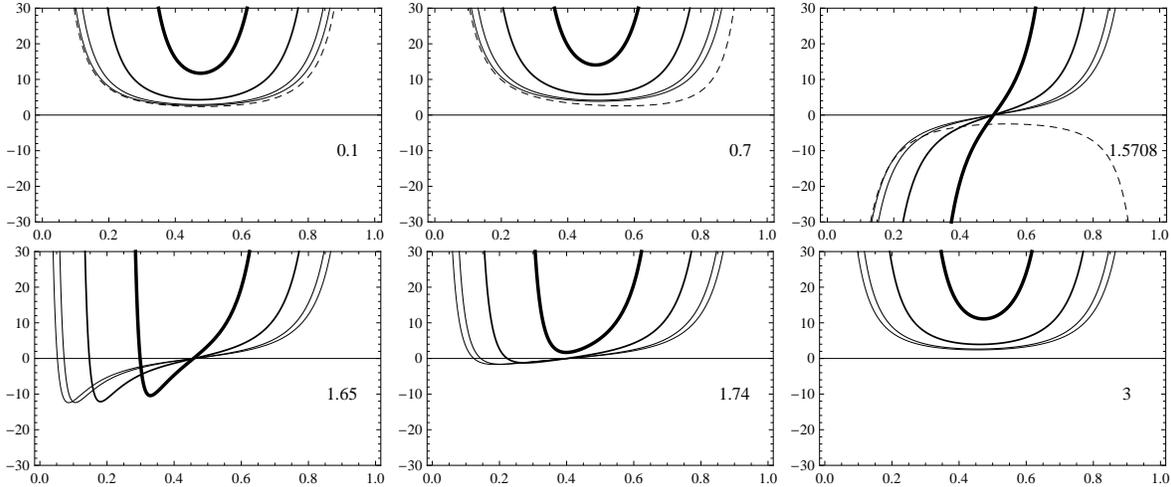}
\caption{Hybrid boundary conditions: DGNN case. Each graph is obtained by fixing the value of the parameter $\alpha$, 
displayed in the lower right corner, in the intervals $[0,\pi/4]$ and $[\pi/2,\pi]$. The parameter $a$ varies along the $x$-axis, 
while the Casimir force on the piston (in units for which $h=c=1$) varies along the $y$-axis. }\label{fig6}
\end{figure}

As a last example we consider Dirichlet boundary conditions on the left end of the piston configuration 
and Neumann boundary condition on the right end. To the left side of the thick piston we impose general boundary conditions and on the right side
of the piston we impose Neumann boundary conditions. This configuration, which we name DGNN, is obtained from (\ref{5})-(\ref{7}) by using the values
\begin{equation}\label{49}
A_{1}=1\;,A_{2}=0\;, B_{1}=\sin\alpha\;, B_{2}=\cos\alpha\;, C_{1}=0\;, C_{2}=1\;, D_{1}=0\;, D_{2}=1\;.
\end{equation} 
Also in this case the parameter $\alpha$ is constrained to belong in the intervals $[0,\pi/4]$ or $[\pi/2,\pi]$ and plots of the Casimir force on a thick piston
are provided in Figure \ref{fig6}. The Casimir force in the limit of infinitely thin piston for this particular example is the same as the one obtained in the 
DGDN case and is represented by the dashed lines.

It is worth, at this point, to briefly comment on the results obtained in this Section for the Casimir force.  

In every example considered here one can notice that the Casimir force acting on the piston increases in absolute value as the thickness of the piston increases. 
In other words, the finite thickness somewhat ``amplifies" the Casimir force acting on the piston.    
There are a few cases in which the Casimir force vanishes at more than one point in the interval $[0,1]$, a 
behavior that has been already observed for infinitely thin pistons with general boundary conditions \cite{fucci15}. It is interesting to notice that 
the position of one of the two equilibrium points varies as the thickness of the piston changes.   
In all of the examples provided above, there are specific values (or a range of values) of the parameter $\alpha$ in which the 
force acting on a thick piston differs substantially from the force acting on the corresponding infinitely thin piston. For instance, in the DGDD
case one can notice that when $\alpha\in[0,\pi/4]$ the force on the infinitely thin piston presents a point of stable equilibrium while for a piston of
finite thickness the force always tend to move it towards the right end of the piston configuration. 
This appears to be a very interesting occurrence and shows that in certain circumstances the Casimir force acting 
on a thick piston behaves quite differently than the force on the corresponding infinitely thin piston.

\section{Concluding Remarks}

In this paper we have analyzed the Casimir energy and force for thick piston configurations endowed 
with general boundary conditions These configurations are modeled by considering a product manifold $[0,L]\times N$ 
and by dividing it into three distinct regions separated by the manifolds $N_{a-\epsilon/2}$ and $N_{a+\epsilon/2}$. The 
thick piston itself, of thickness $\epsilon$, is represented by the region between the cross-sections $N_{a-\epsilon/2}$ and $N_{a+\epsilon/2}$. 
Here we have utilized the spectral zeta function regularization technique in order to compute the Casimir 
energy and the corresponding force. It is important to point out that the results presented in the previous Sections
are very general and are valid for any compact smooth Riemannian base manifold $N$ and any separated boundary 
conditions that lead to a problem with positive eigenvalues.
The explicit expressions for the Casimir energy and force obtained in Sections \ref{sec3} and \ref{sec4} 
have been specialized in the previous Section with the purpose of illustrating a series of particular examples.
For the sake of simplicity, we considered cases in which only the thickness of the piston, $\epsilon$, and one of the parameters describing 
the boundary conditions, $\alpha$, are allowed to vary. It is worth mentioning that due to the general nature of our results 
one could easily construct and analyze examples different than the ones considered here by imposing any combination of allowed boundary conditions and 
by using any allowed thickness $\epsilon$.

The work performed here lends itself to a number of generalizations especially with regard to the geometry of the piston configuration.
For instance it would be interesting to consider thick spherical pistons with general boundary conditions; this analysis would extend to 
thick pistons the results obtained in \cite{dowker11}. An additional straightforward generalization consists in studying thick cylindrical 
pistons with general boundary conditions. The calculations needed to obtain the Casimir energy and force in these cases would follow
the ones presented here with the only difference arising from replacing the sine and cosine functions, obtained as solution to 
(\ref{2}), with the eigenfunctions of the radial equation for the sphere or cylinder. Recently, some work has been focused on
the effect that additional dimensions could have on the Casimir force, see e.g. \cite{cheng06,cheng08,frank07,hoff04,kirsten09}.
It would be interesting to extend the results obtained here to include thick pistons with additional Kaluza-Klein dimensions.
A particularly compelling analysis would be centered on trying to answer the question of how the extra dimensions affect the force on 
pistons of different thickness and general boundary conditions. Perhaps one could find an ideal combination of thickness and boundary conditions that would
somewhat facilitate the detection of possible extra dimensions by measuring the Casimir force on a thick piston.

\end{document}